# The Sixth Generation of the Perseus Digital Library and a Workflow for Open Philology - DRAFT[1]

Gregory Crane, Tufts University, James Tauber, Signum University, Alison Babeu, Lisa Cerrato, Charles Pletcher, Tufts University, Clifford Wulfman, Princeton University, Sergiusz Kazmierski, Regensburg University, Farnoosh Shamsian, Leipzig University.[2]

## Introduction

We report here on the workflow that we needed to develop in order to integrate the growing range of openly licensed, born-digital and, increasingly, machine actionable publications. Our developmental work focused upon textual data for Ancient Greek, Latin, Old English, Classical Arabic and Classical Persian but the challenges that we have had to address are relevant to sources in a wide range of languages, ancient and modern. Perseus 6 was designed to be a publishing workflow that organizes complementary data into an integrated reading environment.[3] This document focuses on the ways in which we have organized the data and describes the current state of ATLAS (Aligned Text and Linguistic Annotation Server) architecture. While this is the sixth version of the Perseus Digital Library, Perseus 6 represents a major step beyond its predecessors. Where Perseus 5 (described below) can represent and integrate digital versions of print editions (e.g., critical editions with interactive textual notes, links to lexicon and commentary entries), Perseus 6 was designed to bring together an expandable range of born-digital classes of annotation. An online [ATLAS server](#) with some initial functionality is now online and public services will expand during the rest of 2024.[4] Most of the ATLAS data is, however, now available on Github and that data will be the focus on this paper in its current version.

The goal of this publication is to introduce people to the problems that we have addressed. Others may build on what we have done or, having seen our work, choose to develop completely different solutions. The work that we have done, however, addresses the need to bring together complementary data sets that are now available under an open license but split across multiple repositories and systems. Our hope is that more people will see what can be done when we bring together the growing wealth of information that open scholarly projects are producing. The work presented here is work in progress and represents, in this draft, development as of November 2024.

---

[1] This draft reflects work that is in rapid development. Much of the work presented was just uploaded to Github in November 2024 and that work will be augmented and refined during November and December of 2024. This paper thus aims to make this work visible as quickly as possible. The paper itself is subject to substantive revision through 2024. Constructive feedback is particularly welcome now.

[2] Gregory Crane was the lead author of this piece. James Tauber was primarily responsible for the new work on the ATLAS architecture performed in 2024.

[3] The work that we describe here was developed as part of the "Perseus on the Web: the Next Thirty Years" project, with support from the US National Endowment for the Humanities, from the Data Intensive Studies Center and from the Faculty of Arts and Sciences at Tufts University. See also (Crane 2019, Crane 2023, Crane et al. 2023, Crane et al. 2024)

[4] Preliminary versions for most ATLAS services are available on an internal Tufts.edu server and are ready to be published on the public-facing https://atlas.perseus.tufts.edu/.

Our goal was to create a workflow to organize, rather than create, textual data that had been produced by, and was available in, platforms that were open but separate. In the quarter century since Creative Commons licenses had emerged in the early 2000s[5], multiple projects in a range of countries had developed robust workflows to produce one or more classes of open data. Projects such as [Perseus](#), [Perseids](#), [Proiel](#), [GLAUx Trees](#), and [Daphne](#) have all published treebanks of Greek and Latin (Hudspeth et al. 2024, Gorman 2020, Keersmaekers et al. 2019, Keersmaekers 2021, Keersmaekers and Van Hal 2022). [Recogito](#) allows users to associate place names in source texts with gazetteers such as [Pleiades](#) and enable automatic mapping (Simon et al 2017, Barker et. al 2024). [INCEpTION](#) (Castilho et al. 2018)enables linguistic and named entity annotation as well as links to authority lists, but we need to turn to separate workflows (such as the [Ugarit Translation Alignment Editor](#) (Palladino et al. 2023) not word and phrase level alignments between source texts and translations). An entirely separate workflow emerged to produce and make available machine actionable metrical analyses for more than 250,000 lines of Greek and Latin poetry (David Chamberlain's [Hypotactic](#)). The [DICES Project](#) publishes metadata identifying and classifying direct speech in Greek and Latin epic (Forstall et al. 2022). Individual scholars publish projects such as SEDES, which identifies which words are in statistically surprising metrical positions in Greek epic (Sansom 2021, Sansom and Fifield 2023). The [Ajax Multi-Commentary Project](#) uses the rich tradition of scholarship on Sophocles to show how we can aggregate and organize multiple commentaries (Romanello and Najem-Meyer 2024). The [Homer Multitext Project](#) publishes new high resolution images and diplomatic editions of, and *scholia* about, the *Iliad*, with links between transcriptions and images (Dué and Ebbott 2019, Smith and Blackwell 2023).

# Background

Before we outline the development of our own work over the past 40 years, we want to articulate some of the principles that have shaped this work and that are relevant to the most recent activities that we describe in the rest of the paper. Sustainable integration of different categories of data has been a driving force behind the development of Perseus from the beginning. Planning for what is now the Perseus Digital Library began in the spring of 1985, with a substantial equipment grant from the Xerox Corporation and a planning grant from the Annenberg/CPB Project. Continuous development began in 1987 and has continued ever since.

The earliest versions of Perseus emphasized two classes of integration. **First**, one inspiration for Perseus was the realization that a single system could display not only textual but also visual information – a fundamentally radical idea in the early 1980s when state of the art computer terminals displayed monowidth (typewriter-style) ASCII characters without any formatting. As a graduate student, co-author Crane needed to move between two different Harvard libraries as he moved from philological to archaeological and art historical data but even the best print publications had relatively few images – typically black and white and very low resolution in comparison with born-digital images. A major goal for Perseus was to combine in a single digital space textual data of various types with visual information such as images, maps, satellite photos, and drawings (Crane 1996, Smith et al. 2000). To demonstrate such integration, we often used the description of Croesus' dedications at Delphi, showing how we could much more effectively bring together maps, images and drawings of the site and of objects than was feasible with print. In the late 1990s as many other projects (the German Arachne Project in particular) began to make digital images and metadata about the art and archaeological record available online, we shifted our focus to the textual record. In 2024, however, we have begun (as will be discussed below) to exploit the International Image Interoperability Framework to begin again integrating the textual and material record.

---

[5] For more on the history of CC licenses, please see: https://creativecommons.org/timeline/

**Second**, already in the 1980s, we exploited automatic analysis to create new links between previously separate classes of textual data. The Morpheus system (Crane 1991) is a rule-based morphological analyzer for Classical Greek and Latin that was first developed in 1985 in the C programming language and still in use.[6] Greek and Latin are morphologically much more complex than languages such as English, French and even German. In print culture, readers have often struggled to match inflected forms on a printed page (e.g., *ênenkas*, "you (sg) carried") with the relevant dictionary entry (e.g., *pherô*, "I carry"). Given an inflected Greek or Latin form, Morpheus provides every possible morphological analysis (e.g., *ênenkas* is 2nd person singular aorist indicative active) and normalized dictionary form (e.g., *pherô*), matching its tables of stems, endings, and combination rules. This system had two basic applications. First, readers could click on a form and then follow links to the morphological analysis and then to machine readable Greek and (starting in the late 1990s) Latin dictionaries. Second, users could generate searches for those forms: e.g., ask for *pherô*, "carry," and retrieve *fereis*, "you (sg.) carry," *eferon*, "I or they were carrying," *oisete*, "you (pl.) will carry," *enêngektai*, "she/he/it has been carried," etc.

**Third**, from the earliest stages of planning, we designed Perseus to be as sustainable as possible. The software development and scholarly explorations that led to Perseus began with work on the *Thesaurus Linguae Graecae*, a corpus of Greek literature then available on magnetic tape for third party development (now locked behind a proprietary paywall). This experience made it clear that content could, and should, be separate from the software for publication and analysis. The lifecycle of software is short while data, especially in fields such as Greco-Roman studies, has value that persists not only over years but over centuries and millennia.

Our work began before the Text Encoding Initiative (TEI) would document guidelines for the use of textual markup, but we already were familiar with the principles behind the TEI. On a snowy night in 1985, the authors of DeRose (1990) presented the case for a generalized model of text content (with a talk that bore the same title as their classic paper: "What is text, really?") As a result we adopted SGML (the predecessor to XML) and would later revise our markup to follow the TEI. The investment in TEI XML was onerous at first – tools for creating and validating SGML were still at an early stage of development and we had to throw out much of the laboriously added markup when we published our sources in HyperCard. The investment would pay dividends over the years as tools improved and the markup increasingly facilitated maintenance and updates. One of Perseus's first major digitization projects, for example, the *Intermediate Liddell Scott Greek-English Lexicon*, was completed in 1985 and is still in active use almost 40 years later. The current version is available on Github

---

[6] Keersmaekers 2019, 2021, 2022.

(https://github.com/helmadik/MiddleLiddell). The GitHub version has, notably, been enhanced by Helma Dik, a scholar who works with, but has never been a part of, the Perseus Project. A sustainability strategy, in our view, should, by the choice of an open license and by public statements, encourage others to take ownership and enhance any digital scholarly product.

## Perseus before Perseus 6.0

The first five versions of the Perseus Digital Library augmented data extracted from print sources (e.g., TEI XML transcriptions of editions and reference works, catalog entries on art objects and archaeological sites converted into metadata) with automatically generated annotations (such as the Morpheus output described above and named entity annotations classifying people, places and organizations and then linking these to authority lists such as the [Pleiades Gazetteer](#)).

- 1992: Perseus 1.0 *Interactive sources and studies on ancient Greece*, published by Yale University Press. This included a Videodisc, CD ROM, and print documentation. Production language: Apple's HyperCard.

- 1995: A web version of the Perseus Digital Library appears at http://www.perseus.tufts.edu/. David A. Smith was the lead developer and the primary programming language was Perl. As noted below, we were able to produce a web version of Perseus years before the CD ROM-based Perseus 2.0 could be published.

- 1997: Perseus 2.0: *Interactive Sources and Studies on Ancient Greece*, published by Yale University Press. This included five CD ROMs and print documentation. Production language: Apple's HyperCard.

    - 2000: Platform Independent Perseus, published by Yale University Press. This contained the same content as Perseus 2.0 but was accessible both on Macintosh and Windows systems.

- 2000: Perseus on the Web becomes sufficiently well-developed that it is listed as Perseus 3.0.

- 2004: Perseus 4.0 ("the Hopper"). This second Web version of the Perseus Digital Library is available at http://www.perseus.tufts.edu/hopper/. David Mimno was the lead developer and the primary programming language was Java. Active development on Perseus 4.0 ended in 2013 but the system continues to run on a suite of virtual machines, supported by Tufts University. Two decades later, Perseus 4.0 remains (as of November 2024) the most commonly used version of the Perseus Digital Library.

- 2018: Perseus 5.0 ("the Scaife Viewer"[7]). James Tauber was the lead developer. The primary development language was Python. Scaife was based upon the Canonical Text Services text model. Scaife built upon the CapiTainS Software Suite and Guidelines for Citable Texts and allowed Perseus to publish new content. This currently includes 2,669 works in 3,776 editions and translations (1,941 in Greek and 631 in Latin), with 83.8 million words in all languages (40.6 million in Greek, 16.4 million in Latin). Brill adopted Scaife as the platform for all of its scholarly editions (Brill's Scholarly Editions). Users of Scaife who have access to these editions (most of which are available behind a paywall) will see the resemblance with Scaife in the page design. As of November, 2024, two editions (*The Literary History of Medicine* and the *Pez Brothers Correspondence*) are open access. While Brill content may be largely proprietary, the contributions that Brill support has made to the Scaife Software Platform are available on Github. In effect, Scaife represents a collaboration between Perseus, an academic project committed to open scholarship, and Brill (now De Gruyter Brill), a traditional publisher that relies upon proprietary control (although it has begun to support open access as a publication option).

Readers scanning the above list of Perseus versions will notice a chronological anomaly: Perseus 3.0 was published in 1995 – two years before Yale would be able to publish Perseus 2.0. Our earlier investment in structured data (such as TEI SGML/XML) allowed David A. Smith to create an initial version of the Perseus Digital Library as an unfunded side-project. While we had worked with Yale University Press to publish the first editions of Perseus and had, in so doing, hoped to help Yale and other university presses develop the infrastructure to develop digital projects, such a development was premature in the 1990s and partnership with a publisher no longer made sense when the World Wide Web emerged in the early 1990s.

The Scaife Viewer was able to support core features of digital editions, including interactive textual notes, automatic dictionary lookup, and integration of commentary and reference works. Brill, in particular, applied Scaife not only to traditional editions but also to editions of so-called fragmentary works (e.g., Jacoby Online), which document surviving evidence about works that have disappeared. Editions of fragmentary works extract from works that do survive quotations that describe, paraphrase, or explicitly quote from works that are otherwise lost. Fragmentary editions are thus meta-editions, with one edition of a fragmentary work building on dozens or hundreds of other editions.[8]

## The "Beyond Translation" Project

The Beyond Translation Project (2019-2023: Crane et al. 2019, 2023, 2023a, 2024.) allowed us to develop a prototype for Perseus 6. Our primary goal was to create a reading environment that could integrate and present a much wider range of born-digital annotations than had been feasible in Perseus 1 through 5. Treebanks, for example, contain not only part of speech tagging and regularized dictionary forms for each word in a corpus but also syntactic role and dependency for each word in a sentence. We needed to be able to represent texts not only as texts with annotations but also graphically as trees. We also needed to be able to represent multiple layers of annotations associated with a text. The following examples illustrate several categories of data that we wanted to represent and that required us to integrate data from standoff markup with one or more textual sources.

---

[7] The Scaife Viewer is named for **Ross Scaife**, a pioneer in digital classics who lived the virtues of collaboration and who set an early example in establishing open access and openly licensed data as the standards upon which Digital Classics now depends. The initial release of the Scaife Viewer was on March 15, 2018, the tenth anniversary of his premature passing on March 15, 2008.

[8] The Berlin Brandenburg Academy of Sciences and the Humanities has also experimented with Scaife and has published sections of its Greek edition and German translation of the *De Locis Affectis* of Galen on Scaife: https://scaife.perseus.org/library/urn:cts:greekLit:tlg0057.tlg057/.

**First**, traditional commentaries constitute one well-known document class that requires standoff markup. Print commentaries can present text on the upper part of the page and commentary on the bottom but the text and commentary are nevertheless separate, if parallel, documents. The basic principle is that commentaries quote and comment upon phrases, words and (often) morphemes or other subsets of words. Perseus 4 simply used citations to link texts and commentary: e.g., a reader looking at Thucydides book 1, chapter 33, section 2 would see any comments that were associated with that chunk of text. With Perseus 6, we now can link from spans in the source text to a commentary.

Figure 1: a commentary on the opening of Thucydides' *History of the Peloponnesian War*: link here.

In Figure 1, yellow highlighting identifies words that have comments. The reader has selected *xunegrapse*, "he composed/wrote," and the system displays the comment on this particular word. The ability to associate particular spans of a text with data can be applied to many classes of annotation, not just traditional commentary.

Figure 2: Links to a language specific (Ancient Greek) grammar: link here.

Figure 2 illustrates links from individual words to explanations in a machine-readable grammar. The reading environment uses highlighting to identify words with grammatical annotations and then displays just those grammatical features that appear in the selected passage. When readers select an annotated word, they see the grammatical explanation on the right while all words with the same grammatical feature have an additional layer of highlighting: e.g., *aeide*, "sing!", *teuche*, "it was fashioning," *eteleieto*, "it was brought to completion" have all been tagged as instances of the imperfect of continuance. Links from spans to external data sets allow us to support an open-ended range of annotation classes, of which the following provide some examples.

**Second**, we consider word and phrase level alignments between source texts and translations. We can generate these alignments between Greek and Latin source texts and English translations with a reasonable accuracy (c. 80%) but we can also manually create alignments between source texts and translations. Further, we can generate born-digital alignments designed to illustrate the literal meanings of a source text. And, of course, we can, in a digital reading environment, offer both literal and literary translations so that readers can choose what they wish to see.

Hafez Farsi / English Word Alignment ⌄     Hide unaligned tokens

الا يا ايها الساقي ادر كاسا و ناولها   1.1.1.1     1.1.1.1   Ho Saki , pass around and offer the bowl of love for God

كه عشق آسان نمود اول ولى افتاد   1.1.1.2     1.1.1.2   For the burden of love for God at first on the day of covenant appeared easy , but now

مشكل‌ها   difficulties have occurred .

**Figure 3: Manual alignment of a Persian poem by Hafez with a 19th century English translation: link [here](#).**

In figure 3 Maryam Foradi manually aligned words and phrases of Persian poetry of Hafez with the 1891 English translation by H. Wilberforce Clarke. In the figure above the reader has moused over "Saki" and sees that name highlighted in the Persian original. More importantly, however, the reader can see that the words presented in light red do not have any equivalents in the original Persian. By showing what the translator has added, the annotator has revealed to readers with no knowledge of Persian that a layer of religious allegory has been imposed upon the original (Palladino et al. 2021, Foradi et al 2019)

Translation alignment can be revealing by itself but becomes much more powerful when readers can go beyond the translation and explore the form and function of each word in a source text.

This leads to the **second** class of data: linguistic annotations. Perseus had begun planning to develop rich linguistic annotation in 2001. More than one million words of Greek and of Latin each are available in manually treebanked form, while machine learning allows us to produce automatically generated treebanks for any online Greek or Latin corpora.

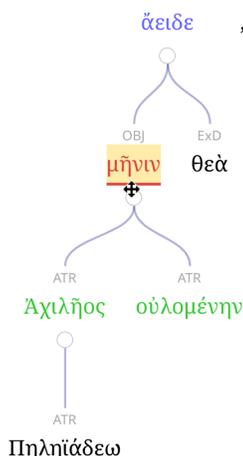

**Figure 4: the opening of the Homeric *Iliad* with linguistic annotations: link here.**

Figure 4 above (left) shows annotations for the *Iliad* that address the needs of readers who wish to push beyond translation alignments and to explore the form and function of each word. There are seven layers of annotation for each word in the opening of the Homeric *Iliad*.

(1) transliteration;
(2) regularized dictionary form;
(3) syntactic role (using a dependency grammar for cross lingual analysis);
(4) part of speech tagging;
(5) a (potentially language specific) grammatical tag;
(6) an English gloss;
(7) a Persian gloss.

These seven layers are drawn from different sources and can, for all practical purposes, only be managed as a series of linked datasets. Layer 1 was automatically generated – unlike languages such as Arabic and Persian, Greek regularly includes short vowels and can be automatically transliterated. Layers 2, 3 and 4 are derived from the Perseus Dependency Treebank. Layer 5 is based on a separate stream of annotations that link words in the text to a language specific Greek grammar to shed light on grammatical features that cannot be inferred from a dependency grammar that was designed to represent shared features across multiple languages. Layers 6 and 7 are glosses into English and Persian. Persian is included because collaborator Farnoosh Shamsian has completed a dissertation on localizing the study of Ancient Greek in Persian.

Translations of the Greek sentence into English and Persian are visible below. The goal is to support an open-ended number of modern languages and to localize the platform as a whole (e.g., replace the English with Persian or some other language).

**Third**, the Greek quoted above is in poetic form and that poetic form is not easy for those studying Greek to decipher. Here we can draw upon yet another class of annotation: machine actionable metrical analyses and recorded readings so that audiences can hear poetry as poetry and learn how to read poetry in metrical form.

ILIAD (GREEK TEXT OF MUNRO & ALLEN)

1.1 μῆ νιν ἄ ει δε θε ά Πη λη ϊ ά δεω Ἀ χι λῆ ος
1.2 οὐ λο μέ νην, ἣ μυ ρί' Ἀ χαι οῖς ἄλ γε' ἔ θη κε,
1.3 πολ λὰς δ'ἰφ θί μους ψυ χὰς Ἄ ϊ δι προ ΐ α ψεν
1.4 ἡ ρώ ων, αὐ τοὺς δὲ ἑ λώ ρι α τεῦ χε κύ νεσ σιν
1.5 οἰ ω νοῖ σί τε πᾶ σι, Δι ὸς δ'ἐ τε λεί ε το βου λή,
1.6 ἐξ οὗ δὴ τὰ πρῶ τα δι α στή την ἐ ρί σαν τε
1.7 Ἀ τρε ΐ δης τε ἄ ναξ ἀν δρῶν καὶ δῖ ος Ἀ χιλ λεύς.

Metrical annotation © 2016 David Chamberlain under CC BY 4.0 License

TEXT WIDTH

AUDIO

© 2016 David Chamberlain under CC BY 4.0 License

DISPLAY MODE
Default
Interlinear
Metrical Annotations

**Figure 5: Metrical analysis and recording of the opening of the *Iliad*: link [here](#).**

The figure above presents a metrical analysis for the opening lines of the *Iliad*, with darker gray representing long syllables, lighter gray shorter syllables and dark vertical bars delineating breaks between the six metrical units of the hexameter. To the right, a button allows readers to listen to a recording of the lines being read. Together the diagram and sound recording make it possible for readers with no Greek to learn how to read Greek poetry (this is now a regular assignment for students who do not know Greek). When readers combine the meter and recording with the aligned translation and linguistic annotation, they are able to engage directly with the Greek in ways that were not before feasible.

Note that the metrical annotations do not correspond with word breaks: metrical units often begin and end in the middle of words. Thus, this class of annotation (along with other forms such as analysis of the morphemes within words) requires an ability to go beyond the token level and annotate chunks of a word.

Other examples discussed elsewhere include the use of named entity annotation to generate maps and social networks and links at the word and character level between transcriptions and images of inscriptions, papyri or manuscripts.

The *Beyond Translation Project* implemented initial front ends for born-digital annotations such as those listed above. In so doing, Beyond Translation also created an initial backend that imported data of different types from different projects into a coherent backend. In late 2023, a new NEH-funded project, *Perseus on the Web: the Next Thirty Years*, reviewed and reorganized the backend in general and the particular formats that we used to make data from multiple sources interoperable.

# Perseus 6, the ATLAS architecture and the CTS Data Model

While Scaife addressed a number of core needs, James Tauber and his collaborators (in particular Jacob Wegner) felt that they needed an approach to complement the Scaife architecture. With funding from Mellon and then from the NEH, they began developing ATLAS, the Aligned Text and Linguistic Annotation Server architecture.

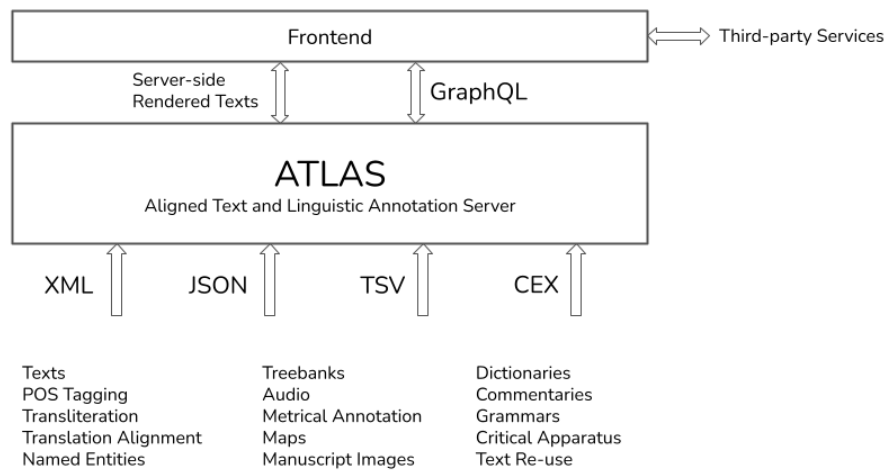

**Figure 6: the ATLAS architecture**

The Canonical Text Services data model allows us to integrate data between CTS-compliant TEI XML and a wider range of data in ATLAS.

The CTS data model allows us to identify chunks of data with a URN such as the following (https://cite-architecture.github.io/ctsurn_spec/):

`urn:cts:greekLit:tlg0012.tlg001.perseus-grc2:1.1-1.7`

- `cts` – This defines the CTS protocol. The data that follows in the URN is unique within this CTS protocol.

- `greekLit` – This is the name space that Perseus and the Open Greek and Latin Project use to define Ancient Greek literature.

- `tlg0012` – This describes a text group. In most cases, a CTS text group will correspond to an author (e.g., Plato or Vergil) but the more general term text group allows us also to deal with cases (such as the New Testament) where we need to be able to address a collection of works by multiple authors as a single unit (`tlg0031`). Likewise, we use `tlg0012` to describe the *Iliad* and the *Odyssey*, whether or not we view these to be the works of a single individual.

- `tlg001` – This describes the particular work and, in this case, `tlg0012.tlg001` designates the Iliad.

- `perseus-grc2` – This designates the particular edition of the *Iliad* that we are citing (in this case, the 1908-1920 Monro/Allen Oxford Classical Text of the Homeric Epics).

- `1.1-1.7` – This defines a text range that extends from line 1 through line 7 of book 1 of the Iliad.

The following sections provide examples for some, but not all, of the annotation classes that we are managing in ATLAS. Readers can follow the repositories https://github.com/scaife-viewer/tagging-pipeline and https://github.com/scaife-viewer/atlas-data-prep for more information.

## Scaife Texts in Atlas

https://github.com/scaife-viewer/tagging-pipeline

While ATLAS helps us integrate data from many different sources, it also provides us with a simpler way to integrate basic textual data. We have completed an initial conversion into ATLAS of all of the texts that are now available in Scaife.

The CapiTainS CTS software library has explicit guidelines for how to structure XML markup, how to organize XML files in a directory, and how to represent metadata. These requirements help develop more interoperable collections and can thus simplify search, analysis and visualization. Nevertheless, the barrier to entry is substantial. ATLAS allows us to add texts that are not CapiTainS-compliant. ATLAS only requires a flat TSV format that includes a unique reference and a chunk of text. We can convert TEI XML into this format, producing a TSV file that contains every book/chapter/section of a particular edition of Thucydides' *History of the Peloponnesian War* or every line from a particular edition of Sophocles' *Antigone*. Thus, the TEI XML version of Henry Stuart Jones edition of Thucydides (available in Scaife) is in the following format:

**Figure 7: opening of Thucydides in TEI XML: link here.**

For Perseus 6, we create a TSV file with the name `tlg0003.tlg001.perseus-grc2.tsv` and then store the lines version of this file for annotation.

```
1    1.1.1    Θουκυδίδης Ἀθηναῖος ξυνέγραψε τὸν πόλεμον τῶν Πελοποννησίων
2    1.1.2    κίνησις γὰρ αὕτη μεγίστη δὴ τοῖς Ἕλλησιν ἐγένετο καὶ μέρει
3    1.1.3    τὰ γὰρ πρὸ αὐτῶν καὶ τὰ ἔτι παλαίτερα σαφῶς μὲν εὑρεῖν διὰ
4    1.2.1    φαίνεται γὰρ ἡ νῦν Ἑλλὰς καλουμένη οὐ πάλαι βεβαίως οἰκουμέ
5    1.2.2    τῆς γὰρ ἐμπορίας οὐκ οὔσης, οὐδ' ἐπιμειγνύντες ἀδεῶς ἀλλήλο
6    1.2.3    μάλιστα δὲ τῆς γῆς ἡ ἀρίστη αἰεὶ τὰς μεταβολὰς τῶν οἰκητόρω
```

**Figure 8: ATLAS TSV version of the opening of Thucydides**: link here.

Because the TSV file becomes the basis for further computation, we can add new texts much more easily by converting them into this TSV identifier+text format. The more heavily structured format of the CapiTainS framework does promote greater consistency, but the TSV identifier+text format is useful for many purposes – indeed, more useful than the XML.

## Morpho-syntactic Analysis

Our goal is to provide multiple layers of linguistic annotation.

**1.** <u>Curated annotations produced by human annotators</u>. More than 2 million words of Greek and Latin have been manually treebanked and made available on GitHub. In at least one case, we have different manual treebanks for the same passage (e.g., both the Proiel Project and Vanessa Gorman have annotated the same passages in Herodotus) and we will need to help people compare different linguistic annotations of comparable authority. For now, however, our goal is to determine whether we have curated annotations for given passages and offer these first.

2. <u>Automatically generated treebanks with some curation</u>. The GLAUx Project has published 20 million words of treebanked Greek, including manually and automatically generated data. The accuracy for the automated work, especially lemmatization and part of speech tagging, is very good. GLAUx includes a number of texts that are not available in the Perseus/Open Greek and Latin collection (particularly short and fragmentary works) but Perseus contains 40 million words of Greek and 16 million words of Latin.

3. <u>Automatically generated treebanks (currently using SpaCy) with the best available models</u>. We use the [greCy](#) and [LatinCy](#) SpaCy pipelines to to generate default morpho-syntactic analyses for all texts in Perseus. The SpaCy treebank data provides a default layer upon which we can rely where we do not have curated treebank data or hybrid human/machine treebanks like GLAUx.

The SpaCy analyses of Open Greek and Latin texts are more bulky (c. 5 gigabytes) than GitHub prefers to support a single repository. We have broken the data up into multiple smaller repositories (the information about how to find data on particular authors is [here](#)). For now, we store this data in the [ConLL-u format](#).

```
1.1.1   0    Θουκυδίδης      PROPN   Ne    Case=Nom|Gender=Masc|Number=Sing                    Θουκυδίδης      nsubj   2
1.1.1   1    Ἀθηναῖος        ADJ     A-    Case=Nom|Degree=Pos|Gender=Masc|Number=Sing         Ἀθηναῖος        amod    0
1.1.1   2    ξυνέγραψε       VERB    V-    Aspect=Perf|Mood=Ind|Number=Sing|Person=3|Tense=Past|VerbForm=Fin|Voice=Act    συγγράφω
1.1.1   3    τὸν             DET     S-    Case=Acc|Definite=Def|Gender=Masc|Number=Sing|PronType=Dem    ὁ       det     4
1.1.1   4    πόλεμον         NOUN    Nb    Case=Acc|Gender=Masc|Number=Sing                    πόλεμος   obj   2
1.1.1   5    τῶν             DET     S-    Case=Gen|Definite=Def|Gender=Masc|Number=Plur|PronType=Dem    ὁ       det     6
1.1.1   6    Πελοποννησίων   ADJ     A-    Case=Gen|Degree=Pos|Gender=Masc|Number=Plur         Πελοποννήσιος   nmod    4
1.1.1   7    καὶ             CCONJ   C-    καί       cc      6
1.1.1   8    Ἀθηναίων        ADJ     A-    Case=Gen|Degree=Pos|Gender=Masc|Number=Plur         Ἀθηναῖος        conj    6
1.1.1   9    ,               PUNCT   Z                                                          ,         dep     2
1.1.1   10   ὡς              SCONJ   G-    ὡς        mark    11
1.1.1   11   ἐπολέμησαν      VERB    V-    Aspect=Perf|Mood=Ind|Number=Plur|Person=3|Tense=Past|VerbForm=Fin|Voice=Act    πολεμέω   ad
1.1.1   12   πρὸς            ADP     R-    πρός      case    13
1.1.1   13   ἀλλήλους        PRON    Pc    Case=Acc|Gender=Masc|Number=Plur|PronType=Rcp       ἀλλήλων   obl     11
```

**Figure 9: Linguistic analysis of the opening of Thucydides using the GreCy pipeline: link [here](#). The figure above shows output for the opening of Thucydides.**

## Dictionaries:

https://github.com/scaife-viewer/atlas-data-prep/tree/main/test-data/dictionaries

We have added most of the dictionaries available in Perseus (notable dictionaries to be added include the *Intermediate Liddell Scott Greek-English Lexicon* and the Slater Pindar Lexicon). The format is JSON and captures the structure and inline formatting in the TEI XML.

An entry from Cunliffe's Lexicon of the Homeric Dialect follows below:
https://github.com/scaife-viewer/atlas-data-prep/tree/main/test-data/dictionaries/cunliffe-1-lex

```
{"headword": "ἀγηνορίη", "data": {"content": "<p>-ης, ἡ</p> <p>[ἀγήνωρ.]</p>"},
"senses":

[{"label": "1", "urn": "urn:cite2:exploreHomer:senses.atlas_v1:1.117",
"definition": "Courage, spirit",

"citations":
```

```
[{"urn": "urn:cite2:scholarlyEditions:citations.v1:1.117_1",
"data": {"ref": "Il. 12.46", "quote": null, "urn":
"urn:cts:greekLit:tlg0012.tlg001.perseus-grc2:12.46"}},
{"urn": "urn:cite2:scholarlyEditions:citations.v1:1.117_2",
"data": {"ref": "Il. 22.457", "quote": null, "urn":
"urn:cts:greekLit:tlg0012.tlg001.perseus-grc2:22.457"}}]},

{"label": "2", "urn": "urn:cite2:exploreHomer:senses.atlas_v1:1.118",
"definition": "The quality in excess or with arrogance.",

"citations": [], "children":
[{"label": "", "urn": "urn:cite2:exploreHomer:senses.atlas_v1:1.119",

"definition": "In pl.", "citations":
[{"urn": "urn:cite2:scholarlyEditions:citations.v1:1.119_1", "data": {"ref": "Il.
9.700", "quote": null, "urn":
"urn:cts:greekLit:tlg0012.tlg001.perseus-grc2:9.700"}}]}]}], "urn":
"urn:cite2:exploreHomer:entries.atlas_v1:1.60"}
```

**Figure 10: An entry from Cunliffe's *Lexicon of the Homeric Dialect*: link [here](here).**

## Textual Notes

https://github.com/scaife-viewer/atlas-data-prep/blob/main/test-data/annotations/commentaries

For now, we store textual notes as a special type of commentary.

```json
{
  "references": [
    "urn:cts:engLit:mds822-32.tpsthl-1599.pdl-eng:1.1"
  ],
  "commentary": "<span>If thou wilt live</span>",
  "fragment": "live with mee",
  "ve_refs": [
    "1.1.t2",
    "1.1.t3",
    "1.1.t4"
  ],
  "idx": "1",
  "urn": "urn:cite2:scaife-viewer:commentary.v1:commentary2",
  "witnesses": [
    {
      "value": "Rs",
      "label": "MS Rodenbach"
    }
  ]
},
```

**Figure 11: A textual note for a poem by Christopher Marlowe.**

Figure 11 above modifies the first line of the *Passionate Shepherd to his Love* by Christopher Marlowe. The default line is: "Come live with mee and be my love." The annotation above indicates that there is a note associated with the phrase "live with me" and that this note, in this case, is a textual variant.

## Text Alignments

https://github.com/scaife-viewer/atlas-data-prep/blob/main/test-data/annotations/text-alignments

```
{
  "urn": "urn:cite2:scaife-viewer:alignment-record.v1:iliad-word-alignment-parrish-998078bc3bab42978b47fa8e8b852cae_3",
  "relations": [
    [
      "urn:cts:greekLit:tlg0012.tlg001.parrish-eng1:1.1.t4",
      "urn:cts:greekLit:tlg0012.tlg001.parrish-eng1:1.1.t5"
    ],
    [
      "urn:cts:greekLit:tlg0012.tlg001.perseus-grc2:1.1.t1"
    ]
  ]
},
```

**Figure 12: A textual alignment between two words in an English translation and one word in a Greek edition of the *Iliad*.**

In figure 12, each alignment between one text and another is a unique annotation with a unique, citable identifier. The format allows for many-to-many alignments. In the example above, for example, two tokens in Amelia Parrish's translation of the first book of the *Iliad* are aligned with one token of Greek.

An individual could create alignments showing how different translations analyzed the same word or the individual could align a pre-existing translation with the source text or they could create a born-digital translation designed for alignment. We use the term text alignments because the same method could be used to align different editions of a Greek text as well as a Greek text with an English or Persian translation.

## Syntax Trees (Treebanks)

https://github.com/scaife-viewer/atlas-data-prep/blob/main/test-data/annotations/syntax-trees/gorman_syntax_trees_017_tlg0008.tlg001.perseus-grc1.json

```
[
  {
    "urn": "urn:cite2:beyond-translation:syntaxTree.atlas_v1:tlg0008-tlg001-grc-1",
    "treebank_id": "1",
    "words": [
      {
        "id": 1,
        "value": "Φύλαρχος",
        "head_id": 79,
        "relation": "SBJ",
        "lemma": "Φύλαρχος",
        "tag": "n-s---mn-"
      },
      {
        "id": 2,
        "value": "δ'",
        "head_id": 79,
        "relation": "AuxY",
        "lemma": "δέ",
        "tag": "d--------"
      }
```

**Figure 13: A treebank represented as JSON**

Figure 13 shows part of a treebank represented as JSON. The tagset for this Treebank is based on the [Perseus Dependency Treebanks](#). Our goal is to begin using the tagset from the [Universal Dependency Framework](#) (UD). That transition can be largely done automatically but will require some curation. The UD data can, however, be represented by the JSON that we have chosen so that we can manage treebanks in multiple formats.

## Audio-annotations

https://github.com/scaife-viewer/atlas-data-prep/blob/main/test-data/annotations/audio-annotations

| urn:cts:greekLit:tlg0012.tlg001.perseus-grc2:1.1 | https://storage.googleapis.com/explorehomer-prod-media/tlg0012.tlg001/audio/1/line_1.mp4 |
| --- | --- |
| urn:cts:greekLit:tlg0012.tlg001.perseus-grc2:1.2 | https://storage.googleapis.com/explorehomer-prod-media/tlg0012.tlg001/audio/1/line_2.mp4 |
| urn:cts:greekLit:tlg0012.tlg001.perseus-grc2:1.3 | https://storage.googleapis.com/explorehomer-prod-media/tlg0012.tlg001/audio/1/line_3.mp4 |
| urn:cts:greekLit:tlg0012.tlg001.perseus-grc2:1.4 | https://storage.googleapis.com/explorehomer-prod-media/tlg0012.tlg001/audio/1/line_4.mp4 |
| urn:cts:greekLit:tlg0012.tlg001.perseus-grc2:1.5 | https://storage.googleapis.com/explorehomer-prod-media/tlg0012.tlg001/audio/1/line_5.mp4 |

**Figure 14: Lines of the Iliad signed to recorded performances of each line.**

At present, we support alignments of text chunks to particular mp4 files. Each line in the TSV file points to a line of the *Iliad*, with the first field pointing to the text of a particular edition and the second to a recorded reading stored in a server.

## Attributions/Credits

https://github.com/scaife-viewer/atlas-data-prep/tree/main/test-data/annotations/attributions

Arguably the most important challenge that we face was to preserve and to aggregate fine-grained credits for born-digital annotations. Credits are easily represented for articles, editions, and even individual commentary notes. In a true digital library, however, a researcher may contribute one or more machine actionable annotations: e.g., they may provide the syntactic structure of a sentence or identify which Antonius is which in a given context or create alignments between a Greek word and its translations in a half dozen passages or may publish metrical analyses for 250,000 lines of Greek and Latin poetry.

In the original workflow for the Perseus Dependency Treebanks, for example, two independent annotators analyzed each sentence while a senior editor reviewed and adjudicated places where the annotators differed. A final corrected version was published, with three credits for each sentence. When another project, run by a former Perseus project member and long-time collaborator, adopted the treebank to a new format, the credits were initially lost. There was no ill intention. There was simply no existing framework to preserve credits in the new format and the project needed to change the format. Likewise, another major treebank project includes both automatically generated and manually curated treebank data. This second project indicates whether a sentence was automatically or manually produced – but the treebank does not identify the annotator. Because the manually produced treebanks are all available on GitHlub, it is possible to identify who did what but this second treebank project did not do so. Its focus was to aggregate existing and produce new treebanks as quickly as possible. Both treebank projects were working with limited time and probably intended to go back to resolve the credits issue but did not find that practical.

In Beyond Translation, we preserved all the credit information that we received. We now have an initial framework by which to represent credits from multiple projects and to make it possible for contributors to develop portfolios showing their contributions across projects.

```
[
  {
    "role": "Annotator",
    "person": {
      "name": "Alex Lessie"
    },
    "organization": {
      "name": "University of Pennsylvania, Philadelphia, PA, USA"
    },
    "data": {
      "references": [
        "urn:cite2:exploreHomer:syntaxTree.v1:syntaxTree-tlg0012-tlg001-perseus-grc2-2277120",
        "urn:cite2:exploreHomer:syntaxTree.v1:syntaxTree-tlg0012-tlg001-perseus-grc2-2277121",
        "urn:cite2:exploreHomer:syntaxTree.v1:syntaxTree-tlg0012-tlg001-perseus-grc2-2277122",
        "urn:cite2:exploreHomer:syntaxTree.v1:syntaxTree-tlg0012-tlg001-perseus-grc2-2277123",
        "urn:cite2:exploreHomer:syntaxTree.v1:syntaxTree-tlg0012-tlg001-perseus-grc2-2277124",
        "urn:cite2:exploreHomer:syntaxTree.v1:syntaxTree-tlg0012-tlg001-perseus-grc2-2277125",
        "urn:cite2:exploreHomer:syntaxTree.v1:syntaxTree-tlg0012-tlg001-perseus-grc2-2277126",
        "urn:cite2:exploreHomer:syntaxTree.v1:syntaxTree-tlg0012-tlg001-perseus-grc2-2277127",
```

```json
[
  {
    "person": {
      "name": "Farnoosh Shamsian"
    },
    "role": "Annotator",
    "organization": {
      "name": "Universität Leipzig: Leipzig, Sachsen, DE"
    },
    "data": {
      "references": [
        "urn:cite2:scaife-viewer:alignment.v1:crito-greek-english-word-alignment-7b34509f15734bd7a20b873aeb08eaa1",
        "urn:cite2:scaife-viewer:alignment.v1:crito-greek-farsi-word-alignment-tr1-7b34509f15734bd7a20b873aeb08eaa1",
        "urn:cite2:scaife-viewer:alignment.v1:crito-greek-farsi-word-alignment-tr2-7b34509f15734bd7a20b873aeb08eaa1",
```

**Figure 15: Attribution data for treebanks and translation alignments.**

Figure 15 shows credits for treebanking individual sentences in the *Iliad* (above) and for aligning particular words and/or phrases between the Greek text of the *Crito* and an English translation and between a Greek text and a Persian translation. We can now begin to aggregate credits:

| Translator | Farshid Rahimi | 3 |
| --- | --- | --- |
| Translator | Mahdi Shojaian | 3 |
| Annotator | Alex Lessie, University of Pennsylvania, Philadelphia, PA, USA | 2,081 |
| Annotator | Daniel Lim Libatique, College of the Holy Cross, Worcester, MA, USA | 293 |
| Annotator | Florin Leonte, Central European University of Budapest, Hungary | 89 |
| Annotator | Jack Mitchell, Tufts University, Medford, MA, USA | 2,410 |

**Figure 16: An initial report showing what contributions individuals have made to the ATLAS server**.

For now, the data is relatively simple. In figure 16, we can see that Farshid Rahimi and Mahdi Shojaian contributed two translated sentences. Alex Lessie helped treebank 2,081 sentences. But the goal is to provide richer information (e.g., allow viewers to see the sentences translated or treebanked) and also to show where annotators contribute to more than one project (e.g., treebanks and translations) but the above is a first step in that direction.

## Next Steps

As the ATLAS backend takes shape, our focus will shift to the next stage of work:

1. We need to build out the services available on the evolving ATLAS server: https://atlas.perseus.tufts.edu/.
2. We need to refine the ATLAS data already available on Github.

3. We need to add frontend support, developed in Beyond Translation (and discussed above) into the Scaife architecture. We will implement Perseus 6 by adding the ATLAS backend and Beyond Translation UI widgets to the earlier Scaife architecture.

## Conclusions and Ongoing Work

This paper provides information about the first release of the ATLAS architecture and representative data.

## Select Bibliography

Barker, E., Palladino, C., & Gordin, S. (2024). Digital Approaches to Investigating Space and Place in Classical Studies. *The Classical Review*, *74*(1), 1–19. https://doi.org/10.1017/S0009840X23002858

Castilho, R. E. D., Klie, J., Kumar, N., Boullosa, B., & Gurevych, I. (2018). Linking Text and Knowledge Using the INCEpTION Annotation Platform. *2018 IEEE 14th International Conference on E-Science (e-Science)*, 327–328. https://doi.org/10.1109/eScience.2018.00077

Crane, G. (1996). Building a Digital Library: The Perseus Project as a Case Study in the Humanities. *DL '96: Proceedings of the First ACM International Conference on Digital Libraries*, 3–10. https://dl.acm.org/doi/pdf/10.1145/226931.226932

Crane, G. (2019). Beyond Translation: Language Hacking and Philology · Issue 1.2, Fall 2019. *Harvard Data Science Review*, *1*(2). https://doi.org/https://doi.org/10.1162/99608f92.282ad764

Crane, G. (1991). Generating and Parsing Classical Greek. *Literary and Linguistic Computing*, 6(4), 243–245. https://doi.org/10.1093/llc/6.4.243

Crane, G. (2023). Perseus 6.0: Beyond Translation, a next generation Perseus Digital Library. *Perseus Journal of Data Preservation and Sustainability*. https://pdldatajournal.pubpub.org/pub/el65xygp/release/1

Crane, G., Babeu, A., Cerrato, L. Shamsian, F., Tauber, J., Wegner, J. (2023). Perseus 6.0: Towards a Next Generation Reading Environment for Born-Digital Editions and Corpora. 2023 ACM/IEEE Joint Conference on Digital Libraries (JCDL), Santa Fe, NM, USA: IEEE, 297–98.

https://doi.org/10.1109/JCDL57899.2023.0006

Crane, G., Shamsian, F., Cerrato, L., Babeu, A., Parrish, A., Penagos, C., Tauber, J., & Wegner, J. (2023). Beyond Translation: engaging with foreign languages in a digital library. *International Journal of Digital Libraries*, 14. https://doi.org/https://link.springer.com/article/10.1007/s00799-023-00349-2

Crane, G., Babeu, A., Cerrato, L., Shamsian, F., Wegner, J., Tauber, J., & Pletcher, C. (2024). *Beyond Translation: translation as gateway rather than endpoint* [White Paper on a concluded project]. Tufts University.

DeRose, S. J., Durand, D. G., Mylonas, E., & Renear, A. H. (1990). What is text, really? *Journal of Computing in Higher Education*, *1*(2), 3–26. https://doi.org/10.1007/BF02941632

Dué, C., & Ebbott, M. (2019). The Homer Multitext within the History of Access to Homeric Epic. In M. Berti (Ed.), *Digital Classical Philology: Ancient Greek and Latin in the Digital Revolution* (pp. 239–256). De Gruyter Saur. https://doi.org/10.1515/9783110599572-014

Foradi, M., Kaßel, J., Pein, J., & Crane, G. R. (2019). Multi-Modal Citizen Science: From Disambiguation to Transcription of Classical Literature. *Proceedings of the 30th ACM Conference on Hypertext and Social Media*, 49–53. https://doi.org/10.1145/3342220.3343667

Forstall, C. W., Finkmann, S., & Verhelst, B. (2022). Towards a linked open data resource for direct speech acts in Greek and Latin epic. *Digital Scholarship in the Humanities,* 37(4), 972–981. https://doi.org/10.1093/llc/fqac006

Hudspeth, M., O'Connor, B., & Thompson, L. (2024). Latin Treebanks in Review: An Evaluation of Morphological Tagging Across Time. *Proceedings of the 1st Workshop on Machine Learning for Ancient Languages (ML4AL 2024)* (pp. 203–218). Association for Computational Linguistics. https://aclanthology.org/2024.ml4al-1.21

Keersmaekers, A., Mercelis, W., Swaelens, C., & Van Hal, T. (2019). Creating, Enriching and Valorizing Treebanks of Ancient Greek. *Proceedings of the 18th International Workshop on Treebanks and Linguistic Theories (TLT, SyntaxFest 2019)*, 109–117. https://doi.org/10.18653/v1/W19-7812

Keersmaekers, A. (2021). The GLAUx corpus: methodological issues in designing a long-term, diverse, multi-layered corpus of Ancient Greek. In N. Tahmasebi, A. Jatowt, Y. Xu, S. Hengchen, S. Montariol, & H. Dubossarsky (Eds.), *Proceedings of the 2nd International Workshop on Computational Approaches to Historical Language Change 2021* (pp. 39–50). Association for Computational Linguistics. https://doi.org/10.18653/v1/2021.lchange-1.6

Keersmaekers, A., & Van Hal, T. (2022). Creating a large-scale diachronic corpus resource: automated parsing in the Greek papyri (and beyond) - KU Leuven. *Natural Language Engineering*. https://doi.org/10.1017/S1351324923000384

Palladino, C., Foradi, M., & Yousef, T. (2021). Translation Alignment for Historical Language Learning: A Case Study. *Digital Humanities Quarterly*, 015(3). http://www.digitalhumanities.org/dhq/vol/15/3/000563/000563.html

Palladino, C., Shamsian, F., Yousef, T., Wright, D. J., Ferreira, A. d'Orange, & Reis, M. F. dos. (2023). Translation Alignment for Ancient Greek: Annotation Guidelines and Gold Standards. *Journal of Open Humanities Data*, 9(1). https://doi.org/10.5334/johd.131

Romanello, M., & Najem-Meyer, S. (2024). A Named Entity-Annotated Corpus of 19th Century Classical Commentaries. *Journal of Open Humanities Data*, *10*(1). https://doi.org/10.5334/johd.150

Sansom, S. A. (2021). Sedes as Style in Greek Hexameter: A Computational Approach. *TAPA (Society for Classical Studies)*, *151*(2), 439–467. https://doi.org/10.1353/apa.2021.0017

Sansom, S. A., & Fifield, D. (2023). SEDES: Metrical Position in Greek Hexameter. *Digital Humanities Quarterly*, *17*(2). https://digitalhumanities.org/dhq/vol/17/2/000675/000675.html

Simon, R., Barker, E., Isaksen, L., & De Soto Cañamares, P. (2017). Linked Data Annotation Without the Pointy Brackets: Introducing Recogito 2. *Journal of Map & Geography Libraries*, *13*(1), 111–132. https://doi.org/10.1080/15420353.2017.1307303

Smith, D. A., Rydberg-Cox, J. A., & Crane, G. R. (2000). The Perseus Project: A Digital Library for the Humanities. *Literary and Linguistic Computing*, *15*, 15–26.

Smith, N., & Blackwell, C. (2023). Analytical developments for the Homer Multitext: Palaeography, orthography, morphology, prosody, semantics. *International Journal on Digital Libraries*, 24(3), 179–184. https://doi.org/10.1007/s00799-023-00380-3

Van Hal, T., & Keersmaekrs, A. (2019). *Pedalion Trees*. Pedalion Trees. https://github.com/perseids-publications/pedalion-trees

## Websites Referenced

Ajax Multi-commentary Project: https://github.com/mromanello/ajax-multi-commentary

Ancient Greek and Latin Perseus Dependency Treebanks: https://perseusdl.github.io/treebank_data/

Brill's Scholarly Editions: https://scholarlyeditions.brill.com/

De Gruyter Brill: https://degruyterbrill.com/

CapiTainS Software Suite and Guidelines for Citable Texts: http://capitains.org/).

Daphne: https://github.com/francescomambrini/Daphne

GLAUx Trees: https://glaux.be/

International Image Interoperability Framework: https://iiif.io/

Jacoby Online: https://scholarlyeditions.brill.com/bnjo/.

Perseids Treebanking Environment: https://perseids-publications.github.io/

Perseus 4.0 ("the Hopper"): https://www.perseus.tufts.edu/hopper/

Perseus 5.0 ("the Scaife Viewer): https://scaife.perseus.org/; source code: https://github.com/scaife-viewer,

Perseus – Beyond Translation: https://beyond-translation.perseus.org/

Perseus 6.0 – Scaife Atlas Server: https://atlas.perseus.tufts.edu/

Pleiades: https://pleiades.stoa.org/

Proiel: https://www.hf.uio.no/ifikk/english/research/projects/proiel/

The Literary History of Medicine: https://scholarlyeditions.brill.com/lhom/.
Perseus 2.0: (1997).

The Pez Brothers' Correspondence: https://scholarlyeditions.brill.com/pez/.

SEDES: https://github.com/sasansom/sedes

Ugarit Alignment Editor: https://ugarit.ialigner.com/

Universal Dependency Framework: https://universaldependencies.org/.